\documentclass[aip,amsmath,amssymb,
reprint,%
]{revtex4-1}
\usepackage{graphicx}
\usepackage{dcolumn}
\usepackage{bm}

\usepackage[utf8]{inputenc}
\usepackage[T1]{fontenc}
\usepackage{mathptmx}
\usepackage{etoolbox}

\makeatletter
\def\@email#1#2{%
	\endgroup
	\patchcmd{\titleblock@produce}
	{\frontmatter@RRAPformat}
	{\frontmatter@RRAPformat{\produce@RRAP{*#1\href{mailto:#2}{#2}}}\frontmatter@RRAPformat}
	{}{}
}%
\makeatother
\begin{document}
	
	\preprint{AIP/123-QED}
	
	\title{Measurements of cyclotron resonance of the interfacial states in strong spin-orbit coupled 2D electron gases proximitized with aluminum}
	\author{Prashant Chauhan}
	\affiliation{Department of Physics and Astronomy, The Johns Hopkins University,  Baltimore, Maryland 21218, USA}

	\author{Candice Thomas}
		\affiliation{Department of Physics and Astronomy and Microsoft Quantum Lab West Layfayette, Purdue University, West Lafayette, Indiana 47907 USA}

\author{Tyler Lindemann}
		\affiliation{Department of Physics and Astronomy and Microsoft Quantum Lab West Layfayette, Purdue University, West Lafayette, Indiana 47907 USA}
	
\author{Geoffrey C. Gardner } 
		\affiliation{Department of Physics and Astronomy and Microsoft Quantum Lab West Layfayette, Purdue University, West Lafayette, Indiana 47907 USA}
	
	\author{J.~Gukelberger}
	\affiliation{Microsoft Quantum, One Microsoft Way Redmond, Washington 98052, USA}
	
	\author{M.~J.~Manfra}
		\affiliation{Department of Physics and Astronomy and Microsoft Quantum Lab West Layfayette, Purdue University, West Lafayette, Indiana 47907 USA}
		\affiliation{School of Materials Engineering, and School of Electrical and Computer Engineering, Purdue University, West Lafayette, Indiana 47907, USA}
		
	\author{N.~P.~Armitage}
	\affiliation{Department of Physics and Astronomy, The Johns Hopkins University, Baltimore, Maryland 21218, USA}

	\date{\today}
	
	\begin{abstract}
		Two dimensional electron gasses (2DEG) in InAs quantum wells proximitized by aluminum are promising platforms for topological qubits based on Majorana zero modes. However, there are still substantial uncertainties associated with the nature of the electronic states at the interfaces of these system.  It is challenging to probe the properties of these hybridized states as they are buried under a relatively thick aluminum layer. In this work we have investigated a range of InAs/In$ _{1-\text{x}} $Ga$ _\text{x} $As heterostructures with Al overlayers using high precision time-domain THz spectroscopy.  Despite the thick metallic overlayer, we observe a prominent cyclotron resonance in magnetic field that can be associated with the response of the interfacial states. Measurements of the THz range complex Faraday rotation allow the extraction of the sign and magnitude of the effective mass, density of charge carriers, and scattering times of the 2DEG despite the close proximity of the aluminum layer.   We discuss the extracted band parameters and connect their values to the known physics of these materials.
	\end{abstract}
	
	\maketitle
	
Two dimensional electron gases (2DEG) in the semiconductor quantum well heterostructures have found enormous device applications and are known to be one of the best platforms for exploring new quantum phenomenon~\cite{Hopkins_APL_1991,Ikebe_PRL_2010,Spivak_RMP_2010,shabani2016two}.  This is possible due to the ease of tuning their properties like charge carrier density, mass and mobility. These properties of 2DEGs have been mostly studied using dc transport based techniques like magnetoresistance.  Non-contact probes like IR and Terahertz spectroscopy have been used less often to investigate their properties.  Motivated by theoretical predictions of topological superconductivity in semiconductor–superconductor heterostructures with possible applications in quantum-information processing~\cite{Lutchyn_PRL_2010,Yuval_PRL_2010}, there has been a huge surge in interest in semiconductor-superconducting hybrids systems in the past decade~\cite{Chang_NatNano_2015, Kjaergaard_Natcom_2016, Nichele_PRL_2017}. Some of the more popular hybrids utilize clean elemental metals which are highly conducting as the superconducting component. Examples of this are Al-InAs, Sn-InSb~\cite{Pendharkar_Science_2021}, Ta-InAs~\cite{Damon_AM_2020} and Pb-InAs~\cite{Kanne_Nat_2021,Drachman_PRM_2021}. 2DEG quantum well heterostructures based on semiconductors with large Rashba spin-orbit coefficient like InAs and InSb have been found to be most successful for such applications~\cite{shabani2016two,lutchyn2018majorana}. 

Although the progress in developing these hybrid heterostructures has been impressive, there are still substantial uncertainties associated with the nature of their interfacial states.  Once the highly conducting overlayer is deposited on the semiconducting heterostructure, their coupling can modify the properties of the interface. Contact based probes have the problem of electrical shorting from the highly conducting layer, making it extremely challenging to study the properties of the 2DEG buried below the metallic overlayer~\cite{schuwalow2021band}.   In many cases, properties of the 2DEG can only be studied by removing the metallic overlayer.  Below we detail our development of a measurement scheme to explore the properties of the 2DEG with a highly conducting overlayer using THz-TD magneto-spectroscopy to overcome this challenge.  Such experiments can isolate the cyclotron resonance (CR) response of the 2DEG, as well has having a number of advantages over traditional transport experiments. 

In the present work, we use high precision time-domain magnetoterahertz spectroscopy with polarization modulation in transmission configuration to investigate the CR response in Faraday rotations from the 2DEG in a planar InAs quantum well - aluminum heterostructures.  These experiments, as they measure the frequency and field dependence dependence of a \textit{complex} response function provide a great deal of information. In other contexts they have been able to provide the mass, mobility, and density of multiple conducting channels simultaneously~\cite{cheng2019magnetoterahertz}.  In the present case, we observe a prominent THz frequency CR response from the 2DEG in magnetic fields up to 3 T.    These experiments are sensitive almost exclusively to the 2DEG because of the much smaller mass of the semiconductor layer as compared to that of the elemental metal.  By measuring and modeling the THz range complex Faraday rotation we extract the parameters like the sign and magnitude of the effective mass, density of charge, and scattering times.  Our measurement scheme demonstrates for the first time that it is possible to extract the parameters of a 2DEG that is buried under much more conductive layer by measuring the CR response.

		\begin{figure}
			\includegraphics[width=8.5 cm]{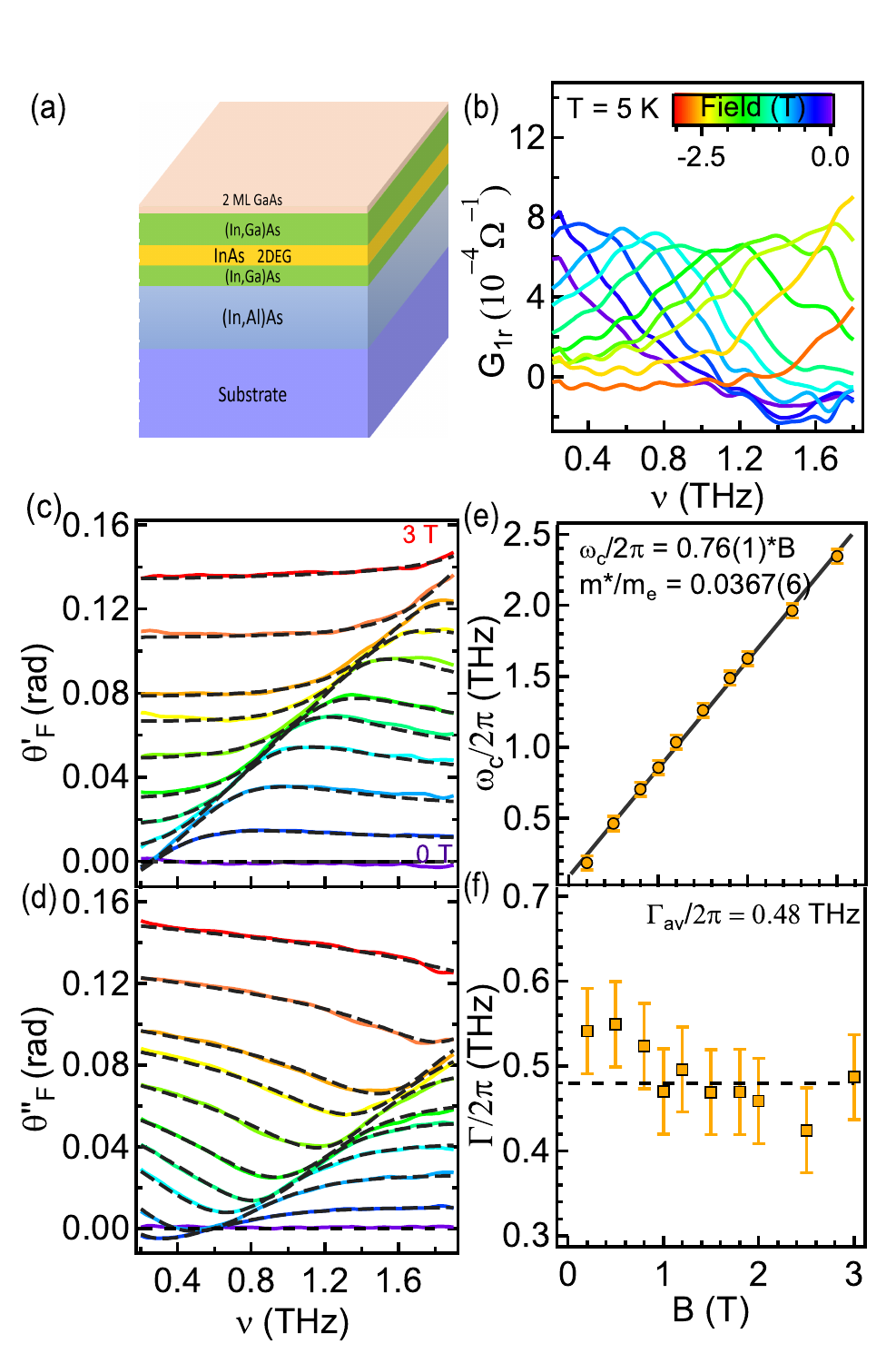}
			\caption{\label{fig:fig1}Full structure without Al. (a) Three-dimensional view of sample geometry with a	cross section of the heterostructure used. The 7 nm wide InAs quantum well (yellow layer) is sandwiched by top 10 nm and bottom 4 nm (In,Ga)As barriers (green) and capped with a 2 ML GaAs layer (upper brown layer). (b) Real part of magneto-optical sheet conductance in circular basis at 5 K for righthand circularly polarized THz light.  Experimental spectra (solid) for (c) real ($ \theta $') and (d) imaginary ($ \theta$'') part of complex Faraday rotation. The dashed lines are fit to the  Faraday rotation as described in the text. The plots are shifted vertically  in proportion to the magnetic field ($ \propto B $) along $ \theta_F $ (e) Extracted field dependent cyclotron resonance frequency $ \omega_c $ with linear fit (solid line). (f) Extracted field dependent scattering rates from the fits along with the average line as guide to the eye (dashed).}
		\end{figure}

\begin{figure}
		\includegraphics[width=8.5 cm]{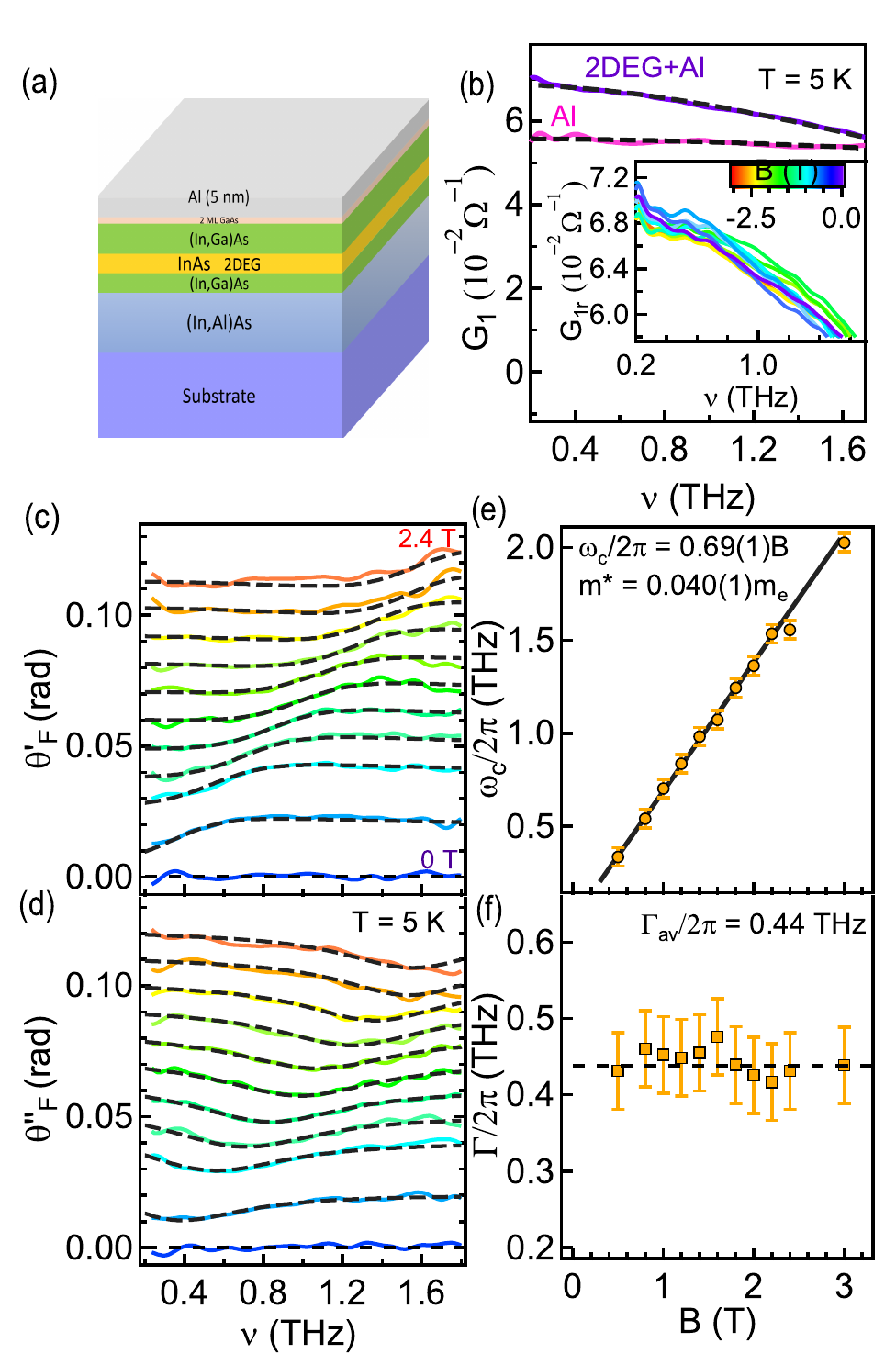}
		\caption{\label{fig:fig2}Full structure with Al overlayer. (a) Three-dimensional view of sample geometry with a	cross section of the heterostructure used. The 7 nm wide InAs quantum well (yellow layer) is sandwiched by top 10 nm and bottom 4 nm (In,Ga)As barriers (green) and capped with a 2 ML GaAs layer (upper brown layer). 7 nm of Al is deposited on top of the capping layer of which top 2 nm is oxidised to insulating AlO$ _\text{x} $ and effectively 5 nm metallic Al is resting on top of the heterostructure.  (b) Real part of sheet conductance in circular basis at 5 K for righthand circularly polarized THz light. Inset: Real part of magneto-optical conductance till 3 T. (c)-(d) Experimental spectra (solid) for real ($ \theta $') and imaginary ($ \theta$'') part of complex Faraday rotation. The dashed lines are fit to the  Faraday rotation as described in the text. The plots are shifted vertically in proportion to the field. (d) Extracted field dependent cyclotron resonance frequency $ \omega_c $ with linear fit (solid line). (e) Extracted field dependent scattering rates from the fits with the average line as guide to the eye (dashed). }
	\end{figure}

The thin film quantum well heterostructures used in this work were grown by molecular beam epitaxy on 0.5 mm thick undoped InP substrate wafer in (100) crystal orientation whose lateral size was $ 1\times1 \text{~cm}^2 $. The main layers of importance in the heterostructure can be seen in the schematic cross section in Fig. \ref{fig:fig1}(a). The 7 nm wide InAs quantum well (yellow layer), is sandwiched by top 10 nm In$_{0.80}$Ga$_{0.20}$As barrier and bottom 4 nm In$_{0.81}$Ga$ _{0.19}$As layer (green) and capped with a 2 ML GaAs layer.  7 nm of Al overlayer was deposited on top;  approximately 2 nm oxidizes to AlO$_\text{x}$ and leaves an effectively 5 nm Al layer on top of the barrier [see Fig. \ref{fig:fig2}(a)]. TDTS experiments were performed in a home-built time-domain terahertz spectrometer with a 7 T magnet in closed cycle cryostat.  On one sample Al was etched off by standard wet etch methods.  Measurements were performed in Faraday geometry meaning both the magnetic field and light propagation direction were same. Both the real and imaginary part of the complex optical conductance $ G(\omega) $ and the Faraday rotation $ \theta_F(\omega) $ were extracted from the measured complex transmission $ T(\omega) $ at 5 K as shown in Refs. \onlinecite{transmission, Bing_APL_2019}. Using the polarization modulation technique~\cite{Morris_OE_2012}, we obtain the complex transmission in right- and left-circularly polarized bases ($ T_r $ and $ T_l $) as it is the transmission eigenbasis for a time-reversal symmetry breaking square system.    We use linearly polarized light and extract out the diagonal and off diagonal components of the Jones transmission matrix and then transform to the right- and left-circularly polarized bases via $T_{r,l} = T_{xx} \pm i T_{xy}$~\cite{armitage2014constraints}.

Fig. \ref{fig:fig1}(b) shows the field dependent real part of the righthand circularly polarized sheet conductance $ G_{1,r}(\omega) $ at 5 K obtained from the complex transmission~\cite{transmission, Bing_APL_2019}. The spectra at zero field show a tail of narrow Drude-like term. At finite field a dip is observed to shift with increasing magnetic field, which is consistent with a CR that depends on field in the usual way $\omega_c = eB/m^*$. The width of the dip given by its full-width at half maximum (FWHM) is a measure of the scattering rate $ \Gamma $, while its area is a measure of the charge density $ n_{2D} /m $.  To get a more accurate measure of these quantities we extract and analyze the complex Faraday rotation. 

We obtain the complex Faraday rotation from the right- and left-hand complex transmission using the expression $ \theta_F(\omega) = -\text{arctan}[i(T_r-T_l)/(T_r+T_l)] $. The $ \theta'_F $ gives the rotation of the major axis of light and the $ \theta''_F $ relates to the ellipticity~\cite{Morris_OE_2012}. Field dependent real ($ \theta'_F $) and imaginary ($ \theta''_F $) parts of the $ \theta_F(\omega) $ up to 3 T are shown in Fig. \ref{fig:fig1}(c)-(d). Note that we have uncertainties $ \theta_F(\omega) $ with field (i.e. $ \theta_F=(\theta_F(B)-\theta_F(-B))/2 $) in order to eliminate a number of systematic errors. An inflection point in $ \theta'_F $ and the dip in $ \theta''_F $ are observed to shift higher in frequency with increasing magnetic field, which is in accordance with the known behavior of CR with $\nu _c = \omega_c / 2 \pi = eB/2 \pi m^* $.

In order to extract the exact CR frequencies and other parameters at different fields, we fit the complex $ \theta_F(\omega) $ to the Drude model. The expression for a single Drude oscillator's magneto-conductance is
\begin{equation}
	G_{r,l}=\epsilon_0 \left[\frac{\omega^2_{pD}d}{\Gamma-i(\omega\pm\omega_c)}-i\omega d(\epsilon_{\infty}-1)\right]. \label{eq1}
\end{equation}
Here, the $ \pm $ sign signifies the use of right- or left-hand circularly polarized light, respectively. $ \omega_{pD} $ is the plasma frequency, d is the film thickness, $ \omega_c $ is the CR frequency, $ \Gamma $ is the scattering rate, and $ \epsilon_{\infty} $ is the background dielectric constant that originates from lattice polarizibility and excitations at frequencies well above the measured spectral range. The spectral weight $\omega^2_{pD}d $ relates to the carrier density ($ n_{2D} $) and the effective transport mass ($ m^* $) as $ \omega^2_{pD}d =  n_{2D}e^2/m^*\epsilon_0 $. The complex Faraday rotation angle can be expressed in terms of right- and left-circularly polarized optical conductance as,
\begin{equation}
	\theta_F(\omega)=\arctan{\left[\frac{i(G_r(\omega)-G_l(\omega))}{G_r(\omega)+G_l(\omega)+2(n+1)/Z_0}\right]}. \label{eq2}
\end{equation}
As shown in Fig. \ref{fig:fig1}(c)-(d), fits to this model while accounting for the field dependence of the Drude term are in good agreement with both the real and imaginary parts of the complex $ \theta_F $. The parameters for the spectral weight $ \omega^2_{pD}d $ and $ d(\epsilon_{\infty}-1) $ are constrained by doing a global fit to the entire field dependent $ \theta_F(\omega) $, leaving the cyclotron frequency $ \omega_c $ and the scattering rate $ \Gamma $ as the only free parameters at each field which allows for their extraction as a function of the magnetic field [see Fig. \ref{fig:fig1}(e)-(f)].   As expected, the cyclotron resonance frequency has linear dependence on the field.  The relation between carrier mass $ m^* $ and resonance frequency (given by $ \omega_c = eB/m^*$~\cite{Liang_PRL_2015}, a fit to the $ \omega_c(B) $ is shown in Fig. \ref{fig:fig1}(e)) gives the cyclotron mass of the electrons in this 2DEG without an Al overlayer to be $0.0367(6)m_e  $, where $ m_e $ is the free electron mass. Then using the relation, $   \omega^2_{pD}d =  n_{2D}e^2/m^*\epsilon_0 $ and spectral weight obtained from fitting the Faraday rotation we extract the total sheet carrier density $ n_{2D} = 3.76\pm0.13 \times10^{11}$ cm$ ^{-2} $. The extracted scattering rate $ \Gamma/2\pi $ of the carriers in field as shown in Fig. \ref{fig:fig1}(f), does not show strong field dependence and has an average of $ \Gamma_{av}/2\pi=0.49(4) $ THz. We determine the mobility of the 2DEG carriers to be $ \mu= e/m^*\Gamma_{av}=1.56\pm0.13\times10^4 $ cm$ ^2/ $V$ \cdot $s.

The above analysis gives us important insight into the properties of the 2DEG in a planar heterostructure.  Some of this information can be inferred from conventional transport experiments, but once a highly conductive overlayer like Al is deposited on top of such a heterostructure (e.g. superconductor-semiconductor devices), it becomes impossible to access the properties of 2DEG using such techniques.   The TDTS CR resonance experiments still provide important information as the effective mass of the aluminum layer is of order the free electron mass, which (when considering the large scattering in the Al layer) is large enough that it does not exhibit a prominent CR.  Therefore the conductance of the Al layer is almost field independent.  As it is a resonance effect, the CR of the 2DEG is still detectable even with transmission greatly decreased because of the Al layer.

We now use the same method of magnetoterahertz spectroscopy to measure the 2DEG on a similar InAs quantum-well heterostructure, but one with an Al overlayer.   Through precise measurements of the complex $ \theta_F$ we can determine the properties of the 2DEG below the metallic overlayer. The schematics of the measured heterostructure with 5 nm Al overlayer is shown in Fig. \ref{fig:fig2}(a). In Fig. \ref{fig:fig2}(b) we show the real part of the zero field sheet conductance $ G_{1}(\omega) $ of both the entire heterostructure as well as a similarly prepared structure that has no 2DEG (e.g. that is just Al).  One can see that the overall scale of the conductance is much larger than that of the 2DEG (Fig. \ref{fig:fig1}(b)).  The conductive layers of the heterostructure (2DEG and Al overlayer) can be treated as parallel conductance channels, so the total conductance will be sum of their individual conductances; thus allowing for larger conductance of the heterostructure in comparison to just Al.  $ G_{1}(\omega) $ for heterostructure with Al is two orders of magnitude larger than the ones without Al [see Fig. \ref{fig:fig1}(b)], showing the large contribution of Al ($ G_{\text{Al}} $) in the effective conductance of the entire heterostructure. The field dependence of the heterostructure conductance up to 3 T is shown in the inset of  Fig. \ref{fig:fig2}(b). A broad feature can be observed to shift with field but the large background of $ G_{\text{Al}} $ makes finding the $ \omega_c $ difficult. In order to better understand the 2DEG properties we look at the in-field Faraday rotation, as $ \theta_F $ response is largely set by the properties of the 2DEG. 

\begin{figure}
	\includegraphics[width=7 cm]{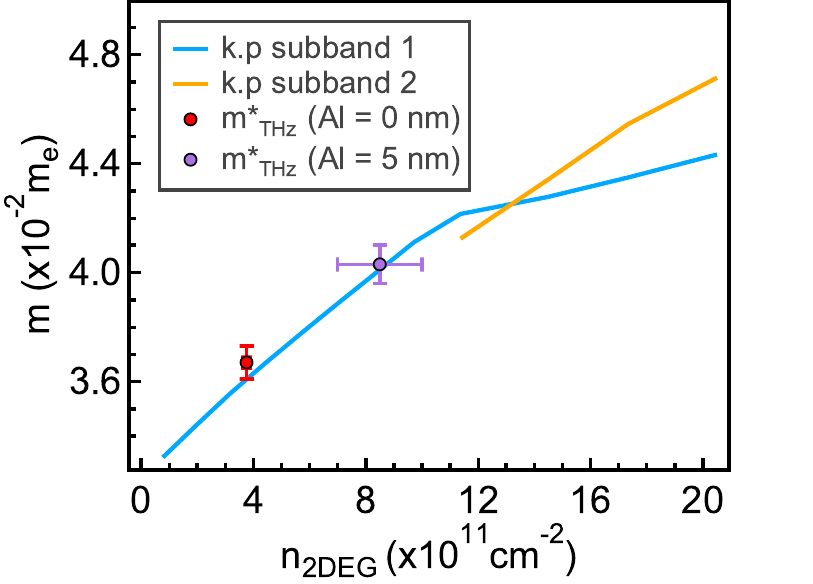}
	\caption{\label{fig:fig3}Mass as a function of density calculated from $\textbf{k} \cdot \textbf{p}$ model for top 25 nm of the semiconductor stack, when the Fermi level ($ E_F $) is pinned close to (80 meV above) the conduction band edge at the  semiconductor/vacuum (semiconductor/aluminum) interface, respectively. The mass and density obtained from the THz measurements of both the samples with and without Al overlayer are compared to the model. }
	\end{figure}

The antisymmetrized complex $ \theta_F(\omega) $ obtained from the complex transmission is shown in Fig. \ref{fig:fig3}(c)-(d). The maximum amplitude of the $ \theta_F(\omega) $  for the sample with Al is $ \sim3 $ times smaller than the sample without Al, implying the suppression of $ \theta_F $ angle in heterostructures with Al. From Eq.(\ref{eq2}), the Faraday rotation depends on the \textit{total} conductance of the sample, which is the sum of conductances of the heterostructure and Al.  The reduction in the rotation angle is due to the Al overlayer as the Al conductance's field dependence is small.  The complex Faraday rotation angle with the Al overlayer is
\begin{equation}
	\theta_F(\omega)=\arctan{\left[\frac{i(G_r(\omega)-G_l(\omega))}{G_r(\omega)+G_l(\omega)+2\left(G_{\text{Al}}+\frac{(n+1)}{Z_0}\right)}\right]} . \label{eq3}
\end{equation}
Here, $ G_{r,l} $ and $ G_{\text{Al}} $ are the conductances for the heterostrusture and the Al overlayer, respectively. Fits to this model of complex $ \theta_F(\omega) $  while constraining the spectral weight $ \omega^2_{pD}d $ and $ d(\epsilon_{\infty}-1) $ by doing a global fit to all the measured fields are shown in Fig. \ref{fig:fig2}(c)-(d). The fits use only a single Drude term and are in reasonably good agreement with the $ \theta_F $ data. The frequency independent contribution from the Al overlayer was accounted for by doing a separate measurement of a 5 nm Al thin film as shown in Fig. \ref{fig:fig2}(b). In Fig. \ref{fig:fig2}(e), the linear fit of $ \omega_c/2\pi$ verses B  gives an effective CR mass of $ 0.0403(7)\text{m}_e $, which is slightly higher than for the sample without Al overlayer. It has larger spectral weight ($ 1.7(3)\times10^-5 $ THz$ ^2\cdot $m) than the sample without Al, giving higher charge density of $ n_{2D}=8.5\pm1.5\times10^{11} $cm$ ^{-2} $. The scattering rate of this sample as shown in Fig. \ref{fig:fig2}(f) is approximately constant with field. The average scattering rate of $ \Gamma_{av}/2\pi =0.44(2)$ THz  gives a mobility of $ \mu=1.57(6)\times10^4 $ cm$ ^2/\text{V}\cdot$s. 

In order to gain further insight into the the 2DEG band structure, we model the quantum well and barrier layers with an 8-band $\textbf{k} \cdot \textbf{p}$ model~\cite{Foreman_PRB_1997,winkler_2003,Vurgaftman_JAP_2001}. The quantum model covers the top 25 nm of the semiconductor stack and is self-consistently coupled to a Poisson solver, which solves for the electrostatic potential between a fixed band offset $ \Phi $ at the surface and a Neumann boundary $\sim$ 1.5 $\mu$m deep in the substrate. By varying the band offset at the surface, we find that the measured densities are reproduced when the Fermi level $ E_F $ is pinned close to (80 meV above) the conduction band edge ($ E_c $) at the semiconductor/vacuum (semiconductor/aluminum) interface, respectively. Both samples (with and without Al) have a single subband, whose density is predominantly located in the InAs quantum well. The subband's cyclotron mass increases monotonically with density [See Fig. \ref{fig:fig3}], as expected for a nonparabolic 2D band where the cyclotron mass $m_c = \pi \hbar^2 \text{DOS}(E_F)$ grows with Fermi energy. The fact that the measured density and mass values lie directly on the predicted curve is a nontrivial consistency check because the model has no free parameters beyond the band offset that was swept to produce the curve.
 
Our measurements give the most accurate determination of parameters of 2DEGs hybridized with aluminum overlayers.   Importantly, our experiments can measure parameters relevant to the 2DEG despite the large conductive background given by the aluminum.  In Table I, we give the fitted parameters for samples without and with the Al overlayer (as well as aluminum to compare).  One can see that the samples with an Al overlayer have a substantially increased density and moderately increased mass.   This can be compared to $\bf{k} \cdot \bf{p}$ calculations that match well the experiment to within experimental uncertainties.   It is important to note that the the agreement is good for the aluminum overlayer samples despite the fact that theoretical calculations do not explicitly include hybridization with aluminum states.  That there is appreciable coupling nevertheless can be seen from that fact that when this structure is cooled down below T$_c$ there will be a large induced gap $\sim200\mu$eV in the semiconductor~\cite{Nichele_PRL_2017}.  One might infer from this that the hybridization is small at least in comparison to the rate set by the measurement frequency. It is also interesting to note that the mobility of these 2DEGs do not appreciably change with etching, which presumably introduces disorder.   This is particularly remarkable because the etched samples have lower density and the effects of disorder on mobility are expected to be larger at low density.  The lack of change may be due to the fact that for the higher density aluminum covered samples, the electron lives in a double well potential with an appreciable part of its wavefunction on the surface.  It is likely that the effect of the wavefunction shifting away from the surface in the etched sample compensates for the larger disorder and its effects at lower densities.

		\begin{table}
	\caption{\label{tab:table4}Properties of the 2DEG in the InAs/In$ _{1-\text{x}} $Ga$ _\text{x} $As heterostructures with and without Al. Note: Aluminum values are taken from~\cite{Ashcroft76,Lin2015-ie} and $ \Phi $ is the conduction band offset ($ E_F-E_c $) used in the $\textbf{k} \cdot \textbf{p}$ calculations. }
	\begin{ruledtabular}
		\begin{tabular}{ccccc}
			System&$ \mbox{m}^*/\mbox{m}_e $&$ \mbox{n} $ ($ 10^{11} $cm$ ^{-2} $)& \mbox{$\mu (10^4$ cm$ ^2/ $V$ \cdot $ s$)$}& $ \Phi$ (meV) \\
			\hline
			InAs&0.0367(6)&$ 3.76(13) $&$ 1.56(13) $& $ 0(15) $\\
			Al-InAs &0.0403(7)& $ 8.5(1.5) $ & $ 1.57(6) $ &$ 80(30) $\\
			Al& 1.4 & $ \sim10^{5} $ &  - &-\\
		\end{tabular}
	\end{ruledtabular}
\end{table}

		\begin{acknowledgments}
		The work at JHU and Purdue was supported by Microsoft Quantum. Our $k \cdot p$ calculations made use of the semicon~\cite{Skolasinski_github} and Kwant~\cite{Kwant_NJP_2014} packages.
		\end{acknowledgments}


\nocite{*}
		\begin{thebibliography}{28}%
			\makeatletter
			\providecommand \@ifxundefined [1]{%
				\@ifx{#1\undefined}
			}%
			\providecommand \@ifnum [1]{%
				\ifnum #1\expandafter \@firstoftwo
				\else \expandafter \@secondoftwo
				\fi
			}%
			\providecommand \@ifx [1]{%
				\ifx #1\expandafter \@firstoftwo
				\else \expandafter \@secondoftwo
				\fi
			}%
			\providecommand \natexlab [1]{#1}%
			\providecommand \enquote  [1]{``#1''}%
			\providecommand \bibnamefont  [1]{#1}%
			\providecommand \bibfnamefont [1]{#1}%
			\providecommand \citenamefont [1]{#1}%
			\providecommand \href@noop [0]{\@secondoftwo}%
			\providecommand \href [0]{\begingroup \@sanitize@url \@href}%
			\providecommand \@href[1]{\@@startlink{#1}\@@href}%
			\providecommand \@@href[1]{\endgroup#1\@@endlink}%
			\providecommand \@sanitize@url [0]{\catcode `\\12\catcode `\$12\catcode
				`\&12\catcode `\#12\catcode `\^12\catcode `\_12\catcode `\%12\relax}%
			\providecommand \@@startlink[1]{}%
			\providecommand \@@endlink[0]{}%
			\providecommand \url  [0]{\begingroup\@sanitize@url \@url }%
			\providecommand \@url [1]{\endgroup\@href {#1}{\urlprefix }}%
			\providecommand \urlprefix  [0]{URL }%
			\providecommand \Eprint [0]{\href }%
			\providecommand \doibase [0]{http://dx.doi.org/}%
			\providecommand \selectlanguage [0]{\@gobble}%
			\providecommand \bibinfo  [0]{\@secondoftwo}%
			\providecommand \bibfield  [0]{\@secondoftwo}%
			\providecommand \translation [1]{[#1]}%
			\providecommand \BibitemOpen [0]{}%
			\providecommand \bibitemStop [0]{}%
			\providecommand \bibitemNoStop [0]{.\EOS\space}%
			\providecommand \EOS [0]{\spacefactor3000\relax}%
			\providecommand \BibitemShut  [1]{\csname bibitem#1\endcsname}%
			\let\auto@bib@innerbib\@empty
			\bibitem [{\citenamefont {Hopkins}\ \emph {et~al.}(1991)\citenamefont
				{Hopkins}, \citenamefont {Rimberg}, \citenamefont {Westervelt}, \citenamefont
				{Tuttle},\ and\ \citenamefont {Kroemer}}]{Hopkins_APL_1991}%
			\BibitemOpen
			\bibfield  {author} {\bibinfo {author} {\bibfnamefont {P.~F.}\ \bibnamefont
					{Hopkins}}, \bibinfo {author} {\bibfnamefont {A.~J.}\ \bibnamefont
					{Rimberg}}, \bibinfo {author} {\bibfnamefont {R.~M.}\ \bibnamefont
					{Westervelt}}, \bibinfo {author} {\bibfnamefont {G.}~\bibnamefont {Tuttle}},
				\ and\ \bibinfo {author} {\bibfnamefont {H.}~\bibnamefont {Kroemer}},\
			}\bibfield  {title} {\enquote {\bibinfo {title} {{Quantum Hall effect in
							InAs/AlSb quantum wells}},}\ }\href {\doibase 10.1063/1.105188} {\bibfield
				{journal} {\bibinfo  {journal} {Applied Physics Letters}\ }\textbf {\bibinfo
					{volume} {58}},\ \bibinfo {pages} {1428--1430} (\bibinfo {year} {1991})},\
			\Eprint {http://arxiv.org/abs/https://doi.org/10.1063/1.105188}
			{https://doi.org/10.1063/1.105188} \BibitemShut {NoStop}%
			\bibitem [{\citenamefont {Ikebe}\ \emph {et~al.}(2010)\citenamefont {Ikebe},
				\citenamefont {Morimoto}, \citenamefont {Masutomi}, \citenamefont {Okamoto},
				\citenamefont {Aoki},\ and\ \citenamefont {Shimano}}]{Ikebe_PRL_2010}%
			\BibitemOpen
			\bibfield  {author} {\bibinfo {author} {\bibfnamefont {Y.}~\bibnamefont
					{Ikebe}}, \bibinfo {author} {\bibfnamefont {T.}~\bibnamefont {Morimoto}},
				\bibinfo {author} {\bibfnamefont {R.}~\bibnamefont {Masutomi}}, \bibinfo
				{author} {\bibfnamefont {T.}~\bibnamefont {Okamoto}}, \bibinfo {author}
				{\bibfnamefont {H.}~\bibnamefont {Aoki}}, \ and\ \bibinfo {author}
				{\bibfnamefont {R.}~\bibnamefont {Shimano}},\ }\bibfield  {title} {\enquote
				{\bibinfo {title} {{Optical Hall Effect in the Integer Quantum Hall
							Regime}},}\ }\href {\doibase 10.1103/PhysRevLett.104.256802} {\bibfield
				{journal} {\bibinfo  {journal} {Phys. Rev. Lett.}\ }\textbf {\bibinfo
					{volume} {104}},\ \bibinfo {pages} {256802} (\bibinfo {year}
				{2010})}\BibitemShut {NoStop}%
			\bibitem [{\citenamefont {Spivak}\ \emph {et~al.}(2010)\citenamefont {Spivak},
				\citenamefont {Kravchenko}, \citenamefont {Kivelson},\ and\ \citenamefont
				{Gao}}]{Spivak_RMP_2010}%
			\BibitemOpen
			\bibfield  {author} {\bibinfo {author} {\bibfnamefont {B.}~\bibnamefont
					{Spivak}}, \bibinfo {author} {\bibfnamefont {S.~V.}\ \bibnamefont
					{Kravchenko}}, \bibinfo {author} {\bibfnamefont {S.~A.}\ \bibnamefont
					{Kivelson}}, \ and\ \bibinfo {author} {\bibfnamefont {X.~P.~A.}\ \bibnamefont
					{Gao}},\ }\bibfield  {title} {\enquote {\bibinfo {title} {Colloquium:
						Transport in strongly correlated two dimensional electron fluids},}\ }\href
			{\doibase 10.1103/RevModPhys.82.1743} {\bibfield  {journal} {\bibinfo
					{journal} {Rev. Mod. Phys.}\ }\textbf {\bibinfo {volume} {82}},\ \bibinfo
				{pages} {1743--1766} (\bibinfo {year} {2010})}\BibitemShut {NoStop}%
			\bibitem [{\citenamefont {Shabani}\ \emph {et~al.}(2016)\citenamefont
				{Shabani}, \citenamefont {Kj{\ae}rgaard}, \citenamefont {Suominen},
				\citenamefont {Kim}, \citenamefont {Nichele}, \citenamefont {Pakrouski},
				\citenamefont {Stankevic}, \citenamefont {Lutchyn}, \citenamefont
				{Krogstrup}, \citenamefont {Feidenhans} \emph {et~al.}}]{shabani2016two}%
			\BibitemOpen
			\bibfield  {author} {\bibinfo {author} {\bibfnamefont {J.}~\bibnamefont
					{Shabani}}, \bibinfo {author} {\bibfnamefont {M.}~\bibnamefont
					{Kj{\ae}rgaard}}, \bibinfo {author} {\bibfnamefont {H.~J.}\ \bibnamefont
					{Suominen}}, \bibinfo {author} {\bibfnamefont {Y.}~\bibnamefont {Kim}},
				\bibinfo {author} {\bibfnamefont {F.}~\bibnamefont {Nichele}}, \bibinfo
				{author} {\bibfnamefont {K.}~\bibnamefont {Pakrouski}}, \bibinfo {author}
				{\bibfnamefont {T.}~\bibnamefont {Stankevic}}, \bibinfo {author}
				{\bibfnamefont {R.~M.}\ \bibnamefont {Lutchyn}}, \bibinfo {author}
				{\bibfnamefont {P.}~\bibnamefont {Krogstrup}}, \bibinfo {author}
				{\bibfnamefont {R.}~\bibnamefont {Feidenhans}},  \emph {et~al.},\ }\bibfield
			{title} {\enquote {\bibinfo {title} {Two-dimensional epitaxial
						superconductor-semiconductor heterostructures: A platform for topological
						superconducting networks},}\ }\href@noop {} {\bibfield  {journal} {\bibinfo
					{journal} {Physical Review B}\ }\textbf {\bibinfo {volume} {93}},\ \bibinfo
				{pages} {155402} (\bibinfo {year} {2016})}\BibitemShut {NoStop}%
			\bibitem [{\citenamefont {Lutchyn}, \citenamefont {Sau},\ and\ \citenamefont
				{Das~Sarma}(2010)}]{Lutchyn_PRL_2010}%
			\BibitemOpen
			\bibfield  {author} {\bibinfo {author} {\bibfnamefont {R.~M.}\ \bibnamefont
					{Lutchyn}}, \bibinfo {author} {\bibfnamefont {J.~D.}\ \bibnamefont {Sau}}, \
				and\ \bibinfo {author} {\bibfnamefont {S.}~\bibnamefont {Das~Sarma}},\
			}\bibfield  {title} {\enquote {\bibinfo {title} {Majorana fermions and a
						topological phase transition in semiconductor-superconductor
						heterostructures},}\ }\href {\doibase 10.1103/PhysRevLett.105.077001}
			{\bibfield  {journal} {\bibinfo  {journal} {Phys. Rev. Lett.}\ }\textbf
				{\bibinfo {volume} {105}},\ \bibinfo {pages} {077001} (\bibinfo {year}
				{2010})}\BibitemShut {NoStop}%
			\bibitem [{\citenamefont {Oreg}, \citenamefont {Refael},\ and\ \citenamefont
				{von Oppen}(2010)}]{Yuval_PRL_2010}%
			\BibitemOpen
			\bibfield  {author} {\bibinfo {author} {\bibfnamefont {Y.}~\bibnamefont
					{Oreg}}, \bibinfo {author} {\bibfnamefont {G.}~\bibnamefont {Refael}}, \ and\
				\bibinfo {author} {\bibfnamefont {F.}~\bibnamefont {von Oppen}},\ }\bibfield
			{title} {\enquote {\bibinfo {title} {{Helical Liquids and Majorana Bound
							States in Quantum Wires}},}\ }\href {\doibase 10.1103/PhysRevLett.105.177002}
			{\bibfield  {journal} {\bibinfo  {journal} {Phys. Rev. Lett.}\ }\textbf
				{\bibinfo {volume} {105}},\ \bibinfo {pages} {177002} (\bibinfo {year}
				{2010})}\BibitemShut {NoStop}%
			\bibitem [{\citenamefont {Chang}\ \emph {et~al.}(2015)\citenamefont {Chang},
				\citenamefont {Albrecht}, \citenamefont {Jespersen}, \citenamefont
				{Kuemmeth}, \citenamefont {Krogstrup}, \citenamefont {Nyg{\aa}rd},\ and\
				\citenamefont {Marcus}}]{Chang_NatNano_2015}%
			\BibitemOpen
			\bibfield  {author} {\bibinfo {author} {\bibfnamefont {W.}~\bibnamefont
					{Chang}}, \bibinfo {author} {\bibfnamefont {S.~M.}\ \bibnamefont {Albrecht}},
				\bibinfo {author} {\bibfnamefont {T.~S.}\ \bibnamefont {Jespersen}}, \bibinfo
				{author} {\bibfnamefont {F.}~\bibnamefont {Kuemmeth}}, \bibinfo {author}
				{\bibfnamefont {P.}~\bibnamefont {Krogstrup}}, \bibinfo {author}
				{\bibfnamefont {J.}~\bibnamefont {Nyg{\aa}rd}}, \ and\ \bibinfo {author}
				{\bibfnamefont {C.~M.}\ \bibnamefont {Marcus}},\ }\bibfield  {title}
			{\enquote {\bibinfo {title} {Hard gap in epitaxial
						semiconductor-superconductor nanowires},}\ }\href {\doibase
				10.1038/nnano.2014.306} {\bibfield  {journal} {\bibinfo  {journal} {Nature
						Nanotechnology}\ }\textbf {\bibinfo {volume} {10}},\ \bibinfo {pages}
				{232--236} (\bibinfo {year} {2015})}\BibitemShut {NoStop}%
			\bibitem [{\citenamefont {Kjaergaard}\ \emph {et~al.}(2016)\citenamefont
				{Kjaergaard}, \citenamefont {Nichele}, \citenamefont {Suominen},
				\citenamefont {Nowak}, \citenamefont {Wimmer}, \citenamefont {Akhmerov},
				\citenamefont {Folk}, \citenamefont {Flensberg}, \citenamefont {Shabani},
				\citenamefont {Palmstr{\o}m},\ and\ \citenamefont
				{Marcus}}]{Kjaergaard_Natcom_2016}%
			\BibitemOpen
			\bibfield  {author} {\bibinfo {author} {\bibfnamefont {M.}~\bibnamefont
					{Kjaergaard}}, \bibinfo {author} {\bibfnamefont {F.}~\bibnamefont {Nichele}},
				\bibinfo {author} {\bibfnamefont {H.~J.}\ \bibnamefont {Suominen}}, \bibinfo
				{author} {\bibfnamefont {M.~P.}\ \bibnamefont {Nowak}}, \bibinfo {author}
				{\bibfnamefont {M.}~\bibnamefont {Wimmer}}, \bibinfo {author} {\bibfnamefont
					{A.~R.}\ \bibnamefont {Akhmerov}}, \bibinfo {author} {\bibfnamefont {J.~A.}\
					\bibnamefont {Folk}}, \bibinfo {author} {\bibfnamefont {K.}~\bibnamefont
					{Flensberg}}, \bibinfo {author} {\bibfnamefont {J.}~\bibnamefont {Shabani}},
				\bibinfo {author} {\bibfnamefont {C.~J.}\ \bibnamefont {Palmstr{\o}m}}, \
				and\ \bibinfo {author} {\bibfnamefont {C.~M.}\ \bibnamefont {Marcus}},\
			}\bibfield  {title} {\enquote {\bibinfo {title} {Quantized conductance
						doubling and hard gap in a two-dimensional semiconductor-superconductor
						heterostructure},}\ }\href {\doibase 10.1038/ncomms12841} {\bibfield
				{journal} {\bibinfo  {journal} {Nature Communications}\ }\textbf {\bibinfo
					{volume} {7}},\ \bibinfo {pages} {12841} (\bibinfo {year}
				{2016})}\BibitemShut {NoStop}%
			\bibitem [{\citenamefont {Nichele}\ \emph {et~al.}(2017)\citenamefont
				{Nichele}, \citenamefont {Drachmann}, \citenamefont {Whiticar}, \citenamefont
				{O'Farrell}, \citenamefont {Suominen}, \citenamefont {Fornieri},
				\citenamefont {Wang}, \citenamefont {Gardner}, \citenamefont {Thomas},
				\citenamefont {Hatke}, \citenamefont {Krogstrup}, \citenamefont {Manfra},
				\citenamefont {Flensberg},\ and\ \citenamefont {Marcus}}]{Nichele_PRL_2017}%
			\BibitemOpen
			\bibfield  {author} {\bibinfo {author} {\bibfnamefont {F.}~\bibnamefont
					{Nichele}}, \bibinfo {author} {\bibfnamefont {A.~C.~C.}\ \bibnamefont
					{Drachmann}}, \bibinfo {author} {\bibfnamefont {A.~M.}\ \bibnamefont
					{Whiticar}}, \bibinfo {author} {\bibfnamefont {E.~C.~T.}\ \bibnamefont
					{O'Farrell}}, \bibinfo {author} {\bibfnamefont {H.~J.}\ \bibnamefont
					{Suominen}}, \bibinfo {author} {\bibfnamefont {A.}~\bibnamefont {Fornieri}},
				\bibinfo {author} {\bibfnamefont {T.}~\bibnamefont {Wang}}, \bibinfo {author}
				{\bibfnamefont {G.~C.}\ \bibnamefont {Gardner}}, \bibinfo {author}
				{\bibfnamefont {C.}~\bibnamefont {Thomas}}, \bibinfo {author} {\bibfnamefont
					{A.~T.}\ \bibnamefont {Hatke}}, \bibinfo {author} {\bibfnamefont
					{P.}~\bibnamefont {Krogstrup}}, \bibinfo {author} {\bibfnamefont {M.~J.}\
					\bibnamefont {Manfra}}, \bibinfo {author} {\bibfnamefont {K.}~\bibnamefont
					{Flensberg}}, \ and\ \bibinfo {author} {\bibfnamefont {C.~M.}\ \bibnamefont
					{Marcus}},\ }\bibfield  {title} {\enquote {\bibinfo {title} {{Scaling of
							Majorana Zero-Bias Conductance Peaks}},}\ }\href {\doibase
				10.1103/PhysRevLett.119.136803} {\bibfield  {journal} {\bibinfo  {journal}
					{Phys. Rev. Lett.}\ }\textbf {\bibinfo {volume} {119}},\ \bibinfo {pages}
				{136803} (\bibinfo {year} {2017})}\BibitemShut {NoStop}%
			\bibitem [{\citenamefont {Pendharkar}\ \emph {et~al.}(2021)\citenamefont
				{Pendharkar}, \citenamefont {Zhang}, \citenamefont {Wu}, \citenamefont
				{Zarassi}, \citenamefont {Zhang}, \citenamefont {Dempsey}, \citenamefont
				{Lee}, \citenamefont {Harrington}, \citenamefont {Badawy}, \citenamefont
				{Gazibegovic}, \citenamefont {het Veld}, \citenamefont {Rossi}, \citenamefont
				{Jung}, \citenamefont {Chen}, \citenamefont {Verheijen}, \citenamefont
				{Hocevar}, \citenamefont {Bakkers}, \citenamefont {Palmstrøm},\ and\
				\citenamefont {Frolov}}]{Pendharkar_Science_2021}%
			\BibitemOpen
			\bibfield  {author} {\bibinfo {author} {\bibfnamefont {M.}~\bibnamefont
					{Pendharkar}}, \bibinfo {author} {\bibfnamefont {B.}~\bibnamefont {Zhang}},
				\bibinfo {author} {\bibfnamefont {H.}~\bibnamefont {Wu}}, \bibinfo {author}
				{\bibfnamefont {A.}~\bibnamefont {Zarassi}}, \bibinfo {author} {\bibfnamefont
					{P.}~\bibnamefont {Zhang}}, \bibinfo {author} {\bibfnamefont {C.~P.}\
					\bibnamefont {Dempsey}}, \bibinfo {author} {\bibfnamefont {J.~S.}\
					\bibnamefont {Lee}}, \bibinfo {author} {\bibfnamefont {S.~D.}\ \bibnamefont
					{Harrington}}, \bibinfo {author} {\bibfnamefont {G.}~\bibnamefont {Badawy}},
				\bibinfo {author} {\bibfnamefont {S.}~\bibnamefont {Gazibegovic}}, \bibinfo
				{author} {\bibfnamefont {R.~L. M.~O.}\ \bibnamefont {het Veld}}, \bibinfo
				{author} {\bibfnamefont {M.}~\bibnamefont {Rossi}}, \bibinfo {author}
				{\bibfnamefont {J.}~\bibnamefont {Jung}}, \bibinfo {author} {\bibfnamefont
					{A.-H.}\ \bibnamefont {Chen}}, \bibinfo {author} {\bibfnamefont {M.~A.}\
					\bibnamefont {Verheijen}}, \bibinfo {author} {\bibfnamefont {M.}~\bibnamefont
					{Hocevar}}, \bibinfo {author} {\bibfnamefont {E.~P. A.~M.}\ \bibnamefont
					{Bakkers}}, \bibinfo {author} {\bibfnamefont {C.~J.}\ \bibnamefont
					{Palmstrøm}}, \ and\ \bibinfo {author} {\bibfnamefont {S.~M.}\ \bibnamefont
					{Frolov}},\ }\bibfield  {title} {\enquote {\bibinfo {title}
					{Parity-preserving and magnetic field-resilient superconductivity in insb
						nanowires with sn shells},}\ }\href {\doibase 10.1126/science.aba5211}
			{\bibfield  {journal} {\bibinfo  {journal} {Science}\ }\textbf {\bibinfo
					{volume} {372}},\ \bibinfo {pages} {508--511} (\bibinfo {year} {2021})},\
			\Eprint
			{http://arxiv.org/abs/https://www.science.org/doi/pdf/10.1126/science.aba5211}
			{https://www.science.org/doi/pdf/10.1126/science.aba5211} \BibitemShut
			{NoStop}%
			\bibitem [{\citenamefont {Carrad}\ \emph {et~al.}(2020)\citenamefont {Carrad},
				\citenamefont {Bjergfelt}, \citenamefont {Kanne}, \citenamefont {Aagesen},
				\citenamefont {Krizek}, \citenamefont {Fiordaliso}, \citenamefont {Johnson},
				\citenamefont {Nygård},\ and\ \citenamefont {Jespersen}}]{Damon_AM_2020}%
			\BibitemOpen
			\bibfield  {author} {\bibinfo {author} {\bibfnamefont {D.~J.}\ \bibnamefont
					{Carrad}}, \bibinfo {author} {\bibfnamefont {M.}~\bibnamefont {Bjergfelt}},
				\bibinfo {author} {\bibfnamefont {T.}~\bibnamefont {Kanne}}, \bibinfo
				{author} {\bibfnamefont {M.}~\bibnamefont {Aagesen}}, \bibinfo {author}
				{\bibfnamefont {F.}~\bibnamefont {Krizek}}, \bibinfo {author} {\bibfnamefont
					{E.~M.}\ \bibnamefont {Fiordaliso}}, \bibinfo {author} {\bibfnamefont
					{E.}~\bibnamefont {Johnson}}, \bibinfo {author} {\bibfnamefont
					{J.}~\bibnamefont {Nygård}}, \ and\ \bibinfo {author} {\bibfnamefont
					{T.~S.}\ \bibnamefont {Jespersen}},\ }\bibfield  {title} {\enquote {\bibinfo
					{title} {Shadow epitaxy for in situ growth of generic
						semiconductor/superconductor hybrids},}\ }\href {\doibase
				https://doi.org/10.1002/adma.201908411} {\bibfield  {journal} {\bibinfo
					{journal} {Advanced Materials}\ }\textbf {\bibinfo {volume} {32}},\ \bibinfo
				{pages} {1908411} (\bibinfo {year} {2020})},\ \Eprint
			{http://arxiv.org/abs/https://onlinelibrary.wiley.com/doi/pdf/10.1002/adma.201908411}
			{https://onlinelibrary.wiley.com/doi/pdf/10.1002/adma.201908411} \BibitemShut
			{NoStop}%
			\bibitem [{\citenamefont {Kanne}\ \emph {et~al.}(2021)\citenamefont {Kanne},
				\citenamefont {Marnauza}, \citenamefont {Olsteins}, \citenamefont {Carrad},
				\citenamefont {Sestoft}, \citenamefont {de~Bruijckere}, \citenamefont {Zeng},
				\citenamefont {Johnson}, \citenamefont {Olsson}, \citenamefont
				{Grove-Rasmussen},\ and\ \citenamefont {Nyg{\aa}rd}}]{Kanne_Nat_2021}%
			\BibitemOpen
			\bibfield  {author} {\bibinfo {author} {\bibfnamefont {T.}~\bibnamefont
					{Kanne}}, \bibinfo {author} {\bibfnamefont {M.}~\bibnamefont {Marnauza}},
				\bibinfo {author} {\bibfnamefont {D.}~\bibnamefont {Olsteins}}, \bibinfo
				{author} {\bibfnamefont {D.~J.}\ \bibnamefont {Carrad}}, \bibinfo {author}
				{\bibfnamefont {J.~E.}\ \bibnamefont {Sestoft}}, \bibinfo {author}
				{\bibfnamefont {J.}~\bibnamefont {de~Bruijckere}}, \bibinfo {author}
				{\bibfnamefont {L.}~\bibnamefont {Zeng}}, \bibinfo {author} {\bibfnamefont
					{E.}~\bibnamefont {Johnson}}, \bibinfo {author} {\bibfnamefont
					{E.}~\bibnamefont {Olsson}}, \bibinfo {author} {\bibfnamefont
					{K.}~\bibnamefont {Grove-Rasmussen}}, \ and\ \bibinfo {author} {\bibfnamefont
					{J.}~\bibnamefont {Nyg{\aa}rd}},\ }\bibfield  {title} {\enquote {\bibinfo
					{title} {Epitaxial pb on inas nanowires for quantum devices},}\ }\href
			{\doibase 10.1038/s41565-021-00900-9} {\bibfield  {journal} {\bibinfo
					{journal} {Nature Nanotechnology}\ }\textbf {\bibinfo {volume} {16}},\
				\bibinfo {pages} {776--781} (\bibinfo {year} {2021})}\BibitemShut {NoStop}%
			\bibitem [{\citenamefont {Drachmann}\ \emph {et~al.}(2021)\citenamefont
				{Drachmann}, \citenamefont {Diaz}, \citenamefont {Thomas}, \citenamefont
				{Suominen}, \citenamefont {Whiticar}, \citenamefont {Fornieri}, \citenamefont
				{Gronin}, \citenamefont {Wang}, \citenamefont {Gardner}, \citenamefont
				{Hamilton}, \citenamefont {Nichele}, \citenamefont {Manfra},\ and\
				\citenamefont {Marcus}}]{Drachman_PRM_2021}%
			\BibitemOpen
			\bibfield  {author} {\bibinfo {author} {\bibfnamefont {A.~C.~C.}\
					\bibnamefont {Drachmann}}, \bibinfo {author} {\bibfnamefont {R.~E.}\
					\bibnamefont {Diaz}}, \bibinfo {author} {\bibfnamefont {C.}~\bibnamefont
					{Thomas}}, \bibinfo {author} {\bibfnamefont {H.~J.}\ \bibnamefont
					{Suominen}}, \bibinfo {author} {\bibfnamefont {A.~M.}\ \bibnamefont
					{Whiticar}}, \bibinfo {author} {\bibfnamefont {A.}~\bibnamefont {Fornieri}},
				\bibinfo {author} {\bibfnamefont {S.}~\bibnamefont {Gronin}}, \bibinfo
				{author} {\bibfnamefont {T.}~\bibnamefont {Wang}}, \bibinfo {author}
				{\bibfnamefont {G.~C.}\ \bibnamefont {Gardner}}, \bibinfo {author}
				{\bibfnamefont {A.~R.}\ \bibnamefont {Hamilton}}, \bibinfo {author}
				{\bibfnamefont {F.}~\bibnamefont {Nichele}}, \bibinfo {author} {\bibfnamefont
					{M.~J.}\ \bibnamefont {Manfra}}, \ and\ \bibinfo {author} {\bibfnamefont
					{C.~M.}\ \bibnamefont {Marcus}},\ }\bibfield  {title} {\enquote {\bibinfo
					{title} {Anodic oxidation of epitaxial superconductor-semiconductor
						hybrids},}\ }\href {\doibase 10.1103/PhysRevMaterials.5.013805} {\bibfield
				{journal} {\bibinfo  {journal} {Phys. Rev. Materials}\ }\textbf {\bibinfo
					{volume} {5}},\ \bibinfo {pages} {013805} (\bibinfo {year}
				{2021})}\BibitemShut {NoStop}%
			\bibitem [{\citenamefont {Lutchyn}\ \emph {et~al.}(2018)\citenamefont
				{Lutchyn}, \citenamefont {Bakkers}, \citenamefont {Kouwenhoven},
				\citenamefont {Krogstrup}, \citenamefont {Marcus},\ and\ \citenamefont
				{Oreg}}]{lutchyn2018majorana}%
			\BibitemOpen
			\bibfield  {author} {\bibinfo {author} {\bibfnamefont {R.~M.}\ \bibnamefont
					{Lutchyn}}, \bibinfo {author} {\bibfnamefont {E.~P.}\ \bibnamefont
					{Bakkers}}, \bibinfo {author} {\bibfnamefont {L.~P.}\ \bibnamefont
					{Kouwenhoven}}, \bibinfo {author} {\bibfnamefont {P.}~\bibnamefont
					{Krogstrup}}, \bibinfo {author} {\bibfnamefont {C.~M.}\ \bibnamefont
					{Marcus}}, \ and\ \bibinfo {author} {\bibfnamefont {Y.}~\bibnamefont
					{Oreg}},\ }\bibfield  {title} {\enquote {\bibinfo {title} {Majorana zero
						modes in superconductor--semiconductor heterostructures},}\ }\href@noop {}
			{\bibfield  {journal} {\bibinfo  {journal} {Nature Reviews Materials}\
				}\textbf {\bibinfo {volume} {3}},\ \bibinfo {pages} {52--68} (\bibinfo {year}
				{2018})}\BibitemShut {NoStop}%
			\bibitem [{\citenamefont {Schuwalow}\ \emph {et~al.}(2021)\citenamefont
				{Schuwalow}, \citenamefont {Schr{\"o}ter}, \citenamefont {Gukelberger},
				\citenamefont {Thomas}, \citenamefont {Strocov}, \citenamefont {Gamble},
				\citenamefont {Chikina}, \citenamefont {Caputo}, \citenamefont {Krieger},
				\citenamefont {Gardner} \emph {et~al.}}]{schuwalow2021band}%
			\BibitemOpen
			\bibfield  {author} {\bibinfo {author} {\bibfnamefont {S.}~\bibnamefont
					{Schuwalow}}, \bibinfo {author} {\bibfnamefont {N.~B.}\ \bibnamefont
					{Schr{\"o}ter}}, \bibinfo {author} {\bibfnamefont {J.}~\bibnamefont
					{Gukelberger}}, \bibinfo {author} {\bibfnamefont {C.}~\bibnamefont {Thomas}},
				\bibinfo {author} {\bibfnamefont {V.}~\bibnamefont {Strocov}}, \bibinfo
				{author} {\bibfnamefont {J.}~\bibnamefont {Gamble}}, \bibinfo {author}
				{\bibfnamefont {A.}~\bibnamefont {Chikina}}, \bibinfo {author} {\bibfnamefont
					{M.}~\bibnamefont {Caputo}}, \bibinfo {author} {\bibfnamefont
					{J.}~\bibnamefont {Krieger}}, \bibinfo {author} {\bibfnamefont {G.~C.}\
					\bibnamefont {Gardner}},  \emph {et~al.},\ }\bibfield  {title} {\enquote
				{\bibinfo {title} {Band structure extraction at hybrid narrow-gap
						semiconductor--metal interfaces},}\ }\href@noop {} {\bibfield  {journal}
				{\bibinfo  {journal} {Advanced Science}\ }\textbf {\bibinfo {volume} {8}},\
				\bibinfo {pages} {2003087} (\bibinfo {year} {2021})}\BibitemShut {NoStop}%
			\bibitem [{\citenamefont {Cheng}\ \emph
				{et~al.}(2019{\natexlab{a}})\citenamefont {Cheng}, \citenamefont {Taylor},
				\citenamefont {Folkes}, \citenamefont {Rong},\ and\ \citenamefont
				{Armitage}}]{cheng2019magnetoterahertz}%
			\BibitemOpen
			\bibfield  {author} {\bibinfo {author} {\bibfnamefont {B.}~\bibnamefont
					{Cheng}}, \bibinfo {author} {\bibfnamefont {P.}~\bibnamefont {Taylor}},
				\bibinfo {author} {\bibfnamefont {P.}~\bibnamefont {Folkes}}, \bibinfo
				{author} {\bibfnamefont {C.}~\bibnamefont {Rong}}, \ and\ \bibinfo {author}
				{\bibfnamefont {N.}~\bibnamefont {Armitage}},\ }\bibfield  {title} {\enquote
				{\bibinfo {title} {Magnetoterahertz response and faraday rotation from
						massive dirac fermions in the topological crystalline insulator pb 0.5 sn 0.5
						te},}\ }\href@noop {} {\bibfield  {journal} {\bibinfo  {journal} {Physical
						review letters}\ }\textbf {\bibinfo {volume} {122}},\ \bibinfo {pages}
				{097401} (\bibinfo {year} {2019}{\natexlab{a}})}\BibitemShut {NoStop}%
			\bibitem [{tra()}]{transmission}%
			\BibitemOpen
			\href@noop {} {}\bibinfo {note} {In the thin film limit, the conductance can
				be extracted using the equation, $ T(\omega) =
				\frac{1+n}{1+n+Z_0G(\omega)}\text{exp}[\frac{i\omega}{2\pi c}(n-1)\Delta L]$.
				Here, $ G(\omega) $ is the complex conductance in the eigenbasis of the
				transmission, $ n $ is the refractive index of the substrate, $ Z_0$ is the
				vacuum impedance, and $ \Delta L $ is the thickness difference between the
				sample and reference substrates.}\BibitemShut {Stop}%
			\bibitem [{\citenamefont {Cheng}\ \emph
				{et~al.}(2019{\natexlab{b}})\citenamefont {Cheng}, \citenamefont {Wang},
				\citenamefont {Barbalas}, \citenamefont {Higo}, \citenamefont {Nakatsuji},\
				and\ \citenamefont {Armitage}}]{Bing_APL_2019}%
			\BibitemOpen
			\bibfield  {author} {\bibinfo {author} {\bibfnamefont {B.}~\bibnamefont
					{Cheng}}, \bibinfo {author} {\bibfnamefont {Y.}~\bibnamefont {Wang}},
				\bibinfo {author} {\bibfnamefont {D.}~\bibnamefont {Barbalas}}, \bibinfo
				{author} {\bibfnamefont {T.}~\bibnamefont {Higo}}, \bibinfo {author}
				{\bibfnamefont {S.}~\bibnamefont {Nakatsuji}}, \ and\ \bibinfo {author}
				{\bibfnamefont {N.~P.}\ \bibnamefont {Armitage}},\ }\bibfield  {title}
			{\enquote {\bibinfo {title} {{Terahertz conductivity of the magnetic Weyl
							semimetal Mn$_3$Sn films}},}\ }\href {\doibase 10.1063/1.5093414} {\bibfield
				{journal} {\bibinfo  {journal} {Applied Physics Letters}\ }\textbf {\bibinfo
					{volume} {115}},\ \bibinfo {pages} {012405} (\bibinfo {year}
				{2019}{\natexlab{b}})},\ \Eprint
			{http://arxiv.org/abs/https://doi.org/10.1063/1.5093414}
			{https://doi.org/10.1063/1.5093414} \BibitemShut {NoStop}%
			\bibitem [{\citenamefont {Morris}\ \emph {et~al.}(2012)\citenamefont {Morris},
				\citenamefont {Aguilar}, \citenamefont {Stier},\ and\ \citenamefont
				{Armitage}}]{Morris_OE_2012}%
			\BibitemOpen
			\bibfield  {author} {\bibinfo {author} {\bibfnamefont {C.~M.}\ \bibnamefont
					{Morris}}, \bibinfo {author} {\bibfnamefont {R.~V.}\ \bibnamefont {Aguilar}},
				\bibinfo {author} {\bibfnamefont {A.~V.}\ \bibnamefont {Stier}}, \ and\
				\bibinfo {author} {\bibfnamefont {N.~P.}\ \bibnamefont {Armitage}},\
			}\bibfield  {title} {\enquote {\bibinfo {title} {Polarization modulation
						time-domain terahertz polarimetry},}\ }\href {\doibase 10.1364/OE.20.012303}
			{\bibfield  {journal} {\bibinfo  {journal} {Opt. Express}\ }\textbf {\bibinfo
					{volume} {20}},\ \bibinfo {pages} {12303--12317} (\bibinfo {year}
				{2012})}\BibitemShut {NoStop}%
			\bibitem [{\citenamefont {Armitage}(2014)}]{armitage2014constraints}%
			\BibitemOpen
			\bibfield  {author} {\bibinfo {author} {\bibfnamefont {N.}~\bibnamefont
					{Armitage}},\ }\bibfield  {title} {\enquote {\bibinfo {title} {{Constraints
							on Jones transmission matrices from time-reversal invariance and discrete
							spatial symmetries}},}\ }\href@noop {} {\bibfield  {journal} {\bibinfo
					{journal} {Physical Review B}\ }\textbf {\bibinfo {volume} {90}},\ \bibinfo
				{pages} {035135} (\bibinfo {year} {2014})}\BibitemShut {NoStop}%
			\bibitem [{\citenamefont {Wu}\ \emph {et~al.}(2015)\citenamefont {Wu},
				\citenamefont {Tse}, \citenamefont {Brahlek}, \citenamefont {Morris},
				\citenamefont {Aguilar}, \citenamefont {Koirala}, \citenamefont {Oh},\ and\
				\citenamefont {Armitage}}]{Liang_PRL_2015}%
			\BibitemOpen
			\bibfield  {author} {\bibinfo {author} {\bibfnamefont {L.}~\bibnamefont
					{Wu}}, \bibinfo {author} {\bibfnamefont {W.-K.}\ \bibnamefont {Tse}},
				\bibinfo {author} {\bibfnamefont {M.}~\bibnamefont {Brahlek}}, \bibinfo
				{author} {\bibfnamefont {C.~M.}\ \bibnamefont {Morris}}, \bibinfo {author}
				{\bibfnamefont {R.~V.}\ \bibnamefont {Aguilar}}, \bibinfo {author}
				{\bibfnamefont {N.}~\bibnamefont {Koirala}}, \bibinfo {author} {\bibfnamefont
					{S.}~\bibnamefont {Oh}}, \ and\ \bibinfo {author} {\bibfnamefont {N.~P.}\
					\bibnamefont {Armitage}},\ }\bibfield  {title} {\enquote {\bibinfo {title}
					{High-resolution faraday rotation and electron-phonon coupling in surface
						states of the bulk-insulating topological insulator
						${\mathrm{cu}}_{0.02}{\mathrm{bi}}_{2}{\mathrm{se}}_{3}$},}\ }\href {\doibase
				10.1103/PhysRevLett.115.217602} {\bibfield  {journal} {\bibinfo  {journal}
					{Phys. Rev. Lett.}\ }\textbf {\bibinfo {volume} {115}},\ \bibinfo {pages}
				{217602} (\bibinfo {year} {2015})}\BibitemShut {NoStop}%
			\bibitem [{\citenamefont {Foreman}(1997)}]{Foreman_PRB_1997}%
			\BibitemOpen
			\bibfield  {author} {\bibinfo {author} {\bibfnamefont {B.~A.}\ \bibnamefont
					{Foreman}},\ }\bibfield  {title} {\enquote {\bibinfo {title} {Elimination of
						spurious solutions from eight-band $\mathbf{k}\ensuremath{\cdot}\mathbf{p}$
						theory},}\ }\href {\doibase 10.1103/PhysRevB.56.R12748} {\bibfield  {journal}
				{\bibinfo  {journal} {Phys. Rev. B}\ }\textbf {\bibinfo {volume} {56}},\
				\bibinfo {pages} {R12748--R12751} (\bibinfo {year} {1997})}\BibitemShut
			{NoStop}%
			\bibitem [{\citenamefont {Winkler}(2003)}]{winkler_2003}%
			\BibitemOpen
			\bibfield  {author} {\bibinfo {author} {\bibfnamefont {R.}~\bibnamefont
					{Winkler}},\ }\href {http://dx.doi.org/10.1007/b13586} {\emph {\bibinfo
					{title} {Spin-orbit coupling effects in two-dimensional electron and Hole
						Systems}}},\ \bibinfo {series} {Springer Tracts in Modern Physics}, Vol.\
			\bibinfo {volume} {191}\ (\bibinfo  {publisher} {Springer},\ \bibinfo {year}
			{2003})\BibitemShut {NoStop}%
			\bibitem [{\citenamefont {Vurgaftman}, \citenamefont {Meyer},\ and\
				\citenamefont {Ram-Mohan}(2001)}]{Vurgaftman_JAP_2001}%
			\BibitemOpen
			\bibfield  {author} {\bibinfo {author} {\bibfnamefont {I.}~\bibnamefont
					{Vurgaftman}}, \bibinfo {author} {\bibfnamefont {J.~R.}\ \bibnamefont
					{Meyer}}, \ and\ \bibinfo {author} {\bibfnamefont {L.~R.}\ \bibnamefont
					{Ram-Mohan}},\ }\bibfield  {title} {\enquote {\bibinfo {title} {Band
						parameters for iii–v compound semiconductors and their alloys},}\ }\href
			{\doibase 10.1063/1.1368156} {\bibfield  {journal} {\bibinfo  {journal}
					{Journal of Applied Physics}\ }\textbf {\bibinfo {volume} {89}},\ \bibinfo
				{pages} {5815--5875} (\bibinfo {year} {2001})},\ \Eprint
			{http://arxiv.org/abs/https://doi.org/10.1063/1.1368156}
			{https://doi.org/10.1063/1.1368156} \BibitemShut {NoStop}%
			\bibitem [{\citenamefont {Ashcroft}\ and\ \citenamefont
				{Mermin}(1976)}]{Ashcroft76}%
			\BibitemOpen
			\bibfield  {author} {\bibinfo {author} {\bibfnamefont {N.~W.}\ \bibnamefont
					{Ashcroft}}\ and\ \bibinfo {author} {\bibfnamefont {N.~D.}\ \bibnamefont
					{Mermin}},\ }\href@noop {} {\emph {\bibinfo {title} {{S}olid {S}tate
						{P}hysics}}}\ (\bibinfo  {publisher} {Holt-Saunders},\ \bibinfo {year}
			{1976})\BibitemShut {NoStop}%
			\bibitem [{\citenamefont {Lin}\ \emph {et~al.}(2015)\citenamefont {Lin},
				\citenamefont {Wu}, \citenamefont {Chang}, \citenamefont {Liang},\ and\
				\citenamefont {Lin}}]{Lin2015-ie}%
			\BibitemOpen
			\bibfield  {author} {\bibinfo {author} {\bibfnamefont {S.-W.}\ \bibnamefont
					{Lin}}, \bibinfo {author} {\bibfnamefont {Y.-H.}\ \bibnamefont {Wu}},
				\bibinfo {author} {\bibfnamefont {L.}~\bibnamefont {Chang}}, \bibinfo
				{author} {\bibfnamefont {C.-T.}\ \bibnamefont {Liang}}, \ and\ \bibinfo
				{author} {\bibfnamefont {S.-D.}\ \bibnamefont {Lin}},\ }\bibfield  {title}
			{\enquote {\bibinfo {title} {Pure electron-electron dephasing in percolative
						aluminum ultrathin film grown by molecular beam epitaxy},}\ }\href@noop {}
			{\bibfield  {journal} {\bibinfo  {journal} {Nanoscale Research Letters}\
				}\textbf {\bibinfo {volume} {10}},\ \bibinfo {pages} {71} (\bibinfo {year}
				{2015})}\BibitemShut {NoStop}%
			\bibitem [{Sko()}]{Skolasinski_github}%
			\BibitemOpen
			\href@noop {} {}\bibinfo {note} {R. Skolasinski et al.,
				https://gitlab.kwant-project.org/semicon/semicon}\BibitemShut {NoStop}%
			\bibitem [{\citenamefont {Groth}\ \emph {et~al.}(2014)\citenamefont {Groth},
				\citenamefont {Wimmer}, \citenamefont {Akhmerov},\ and\ \citenamefont
				{Waintal}}]{Kwant_NJP_2014}%
			\BibitemOpen
			\bibfield  {author} {\bibinfo {author} {\bibfnamefont {C.~W.}\ \bibnamefont
					{Groth}}, \bibinfo {author} {\bibfnamefont {M.}~\bibnamefont {Wimmer}},
				\bibinfo {author} {\bibfnamefont {A.~R.}\ \bibnamefont {Akhmerov}}, \ and\
				\bibinfo {author} {\bibfnamefont {X.}~\bibnamefont {Waintal}},\ }\bibfield
			{title} {\enquote {\bibinfo {title} {Kwant: a software package for quantum
						transport},}\ }\href {\doibase 10.1088/1367-2630/16/6/063065} {\bibfield
				{journal} {\bibinfo  {journal} {New Journal of Physics}\ }\textbf {\bibinfo
					{volume} {16}},\ \bibinfo {pages} {063065} (\bibinfo {year}
				{2014})}\BibitemShut {NoStop}%
		\end{thebibliography}
	\end{document}